\documentclass[aps,preprint,showpacs]{revtex4}
\usepackage{color}
\usepackage{graphicx}
\usepackage{bm}
\begin{document}


%
\title{Possible Quantum Diffusion of Polaronic Muons in Dy$_2$Ti$_2$O$_7$ Spin Ice}
\author{P. Qu\'emerais$^{1,2}$, P. McClarty$^{1,*}$, R. Moessner$^{1}$}
\address{$^1$ Max-Planck-Institut for the Physics of Complex Systems, N\"othnitzer Str. 38, 01187 Dresden, Germany \\
$^2$ Institut N\'eel, CNRS and Universit\'e Joseph Fourier, BP 166, 38042 Grenoble Cedex 9, France}
\email{pmcclarty@pks.mpg.de}
%
\date{\today}
\begin{abstract}
We interpret recent measurements of the zero field muon relaxation rate in the frustrated magnetic pyrochlore Dy$_2$Ti$_2$O$_7$ as resulting from the quantum diffusion of muons in the substance. In this scenario, the plateau observed at low temperature ($<7$ K) in the relaxation rate is due to coherent tunneling of the muons through a spatially disordered spin state and not to any magnetic fluctuations persisting at low temperature. Two further regimes either side of a maximum relaxation rate at $T^* = 50$ K correspond to a crossover between tunnelling and incoherent activated hopping motion of the muon. Our fit of the experimental data is compared with the case of muonium diffusion in KCl.
\end{abstract}

%
\maketitle 

The recent measurement of the zero field $\mu$SR relaxation in Dy$_2$Ti$_2$O$_7$ spin ice (DTO) by  Dunsiger et al. \cite{Dunsiger} joins a long series of puzzling experiments on a diverse range of frustrated magnetic materials over roughly the last fifteen years \cite{Bert,Uemura,Fukaya,Keren2,Hodges,Dunsiger,Gardner,TTOMuon,GSO}. They indicate a relaxation of the spin asymmetry of the muons after they are implanted in the sample in a spin-polarised state. This persists down to the lowest observed temperatures with little temperature dependence in the relaxation rate below a temperature varying from $1-10$ K depending on the compound. 

The interpretation of this relaxation has long been a matter of discussion. At its center lies the question whether the origin of the dynamics in each case is due to intrinsic magnetic fluctuations in the material, or whether the implanted muons are instead more than merely passive probes of the magnetism. 

The latter is a realistic possibility as the muon couples not only weakly to the magnetic degrees of freedom via its spin but also potentially {\em much more} strongly to electric degrees of freedom via its positive charge. As we argue here, this can give rise to new physics interesting on its own right, which is in turn elegantly probed via the magnetic degree of freedom.

In DTO, previous $\mu$SR measurements were made \cite{Bramwell1} and an important debate \cite{Bramwell2,Blundell, Sala} concerning the origin of the muon spin relaxation in spin ice is developing. In spin ices, the moments have an Ising anisotropy and the interactions are frustrated leading to the onset of a highly degenerate spin ice state at low temperatures that is signalled by a heat capacity peak at around $1$ K. The dynamics in this material has been explored using several probes besides $\mu$SR \cite{LagoBlundell,Dunsiger} including susceptibility \cite{Snyder,Matsuhira,Quilliam,Matsuhira2} neutron scattering \cite{Ehlers,Ehlers2,Clancy}, magnetocaloric effect \cite{Orendac}, magnetization relaxation experiments \cite{Matsuhira2}, nuclear forward scattering \cite{Sutter}, NQR \cite{Kitagawa} and three distinct dynamical regimes have been found. Above around $15$ K, the dynamics follows an Arrhenius law controlled by a gap to excited crystal field levels of several hundred Kelvin \cite{Ehlers2}. Between about $1$ K and $15$ K, the dynamics is dominated by tunnelling between magnetic configurations and the temperature dependence is correspondingly weaker than at higher temperatures \cite{Jaubert}. Below about $1$ K, the timescales greatly increase, but there is some evidence for a second Arrhenius regime in a.c. susceptibility \cite{Quilliam} although the moments are static on neutron timescales \cite{Clancy}. Between $2$-$16$ K, large timescales ($\lesssim10^{-3}$ s) have been observed by a.c susceptibility measurements \cite{Snyder}, and between $30$-$90$ K, nuclear forward scattering of synchrotron radiation experiments \cite{Sutter} give characteristic fluctuation times between $10^{-7}$ and $10^{-10}$ s between $30$ and $70$ K respectively.

\begin{figure}
\includegraphics[scale=0.28]{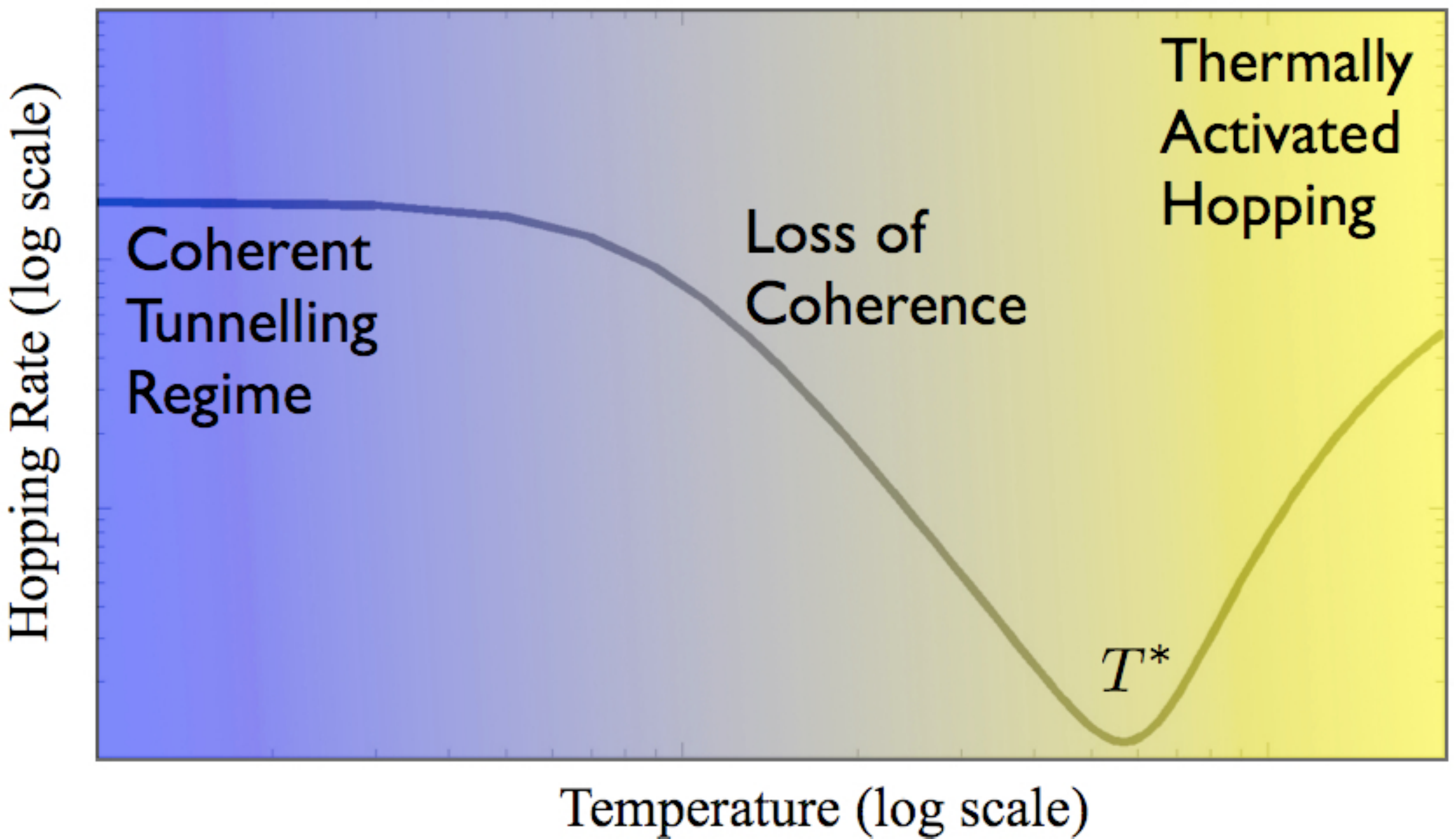}
\caption{\label{fig:colour} Sketch of the temperature variation of the hopping rate of the polaronic muon within the quantum diffusion scenario.}
\end{figure}

The existence of several dynamical regimes appears at odds with the almost featureless $\mu$SR relaxation $1/T_1$ below $7$ K observed in Ref. \cite{Dunsiger} in a zero-field experiment. Moreover, since the moments are, to an excellent approximation, Ising-like there should be insufficient spectral weight at low temperatures to bring about muon spin relaxation. 

In our scenario we assume that the magnetic spins are to a first approximation frozen at low temperatures $T\lesssim 70$ K relative to characteristic times experienced by the muons and that the relaxation mechanism of the muon is due to its diffusion through the static disordered magnetic background\footnote{This approximation is justified because the tunnelling timescale of the muonic polarons is about $10^{-10}$ s. The dominant contribution to muon relaxation is from the shortest timescale \cite{Zaremba,McClarty}. At temperatures higher than about $70$ K, the approximation of static spins is not necessarily justified. At these temperatures, however, both the thermally activated hopping and the intrinsic dynamics (see e.g. \cite{Kitagawa}) are of Arrhenius form with activation energies of the same order of magnitude.}. So far, such an hypothesis has not been considered in insulating oxides. On the contrary, muons are generally believed to be localized in such compounds. However, the diffusion of muons has been observed in many substances including some metals and muonium (a bound state of an electron and a muon, denoted by Mu) diffusion has been seen in some ionic insulators including KCl, NaCl and GaAs (\cite{Kiefl,Kadono,Kadono3,Kadono2}; see \cite{Storchak} for a review). The fact that Mu seems not to have been observed in magnetic insulating oxides such as Dy$_2$Ti$_2$O$_7$ is an indication that its formation is screened by the dielectric constant of the insulator ($\epsilon_s \sim 65$ and $\epsilon_{\infty} \sim 5$ \cite{Bi}, where $\epsilon_s$ and $\epsilon_{\infty}$ are respectively the static and high frequency dielectric constants). Such a high value of the static dielectric constant also indicates that the optical phonons have important effects on the muon: the muon-phonon interactions with both optical and acoustic phonons almost certainly lead to the formation of muonic polarons \cite{Storchak}. 

In the following, we assume the existence of muonic polarons in Dy$_2$Ti$_2$O$_7$ and consider the possibility that these polarons delocalize at low temperatures. In particular, we believe that the diffusion of just such a polaron offers a scenario (Fig.1) within which the data of Dunsiger {\it et al.} \cite{Dunsiger} may be interpreted.

The remainder of this paper is organized as follows.First, we expose how the muon spin relfects the muon diffusion through the disordered spin ice state. Next, we discuss the different regimes of polaron motion, together with a fit to the zero field $\mu$SR data on Dy$_2$Ti$_2$O$_7$. The fitting parameters are then compared to independently obtained estimates and found to be in satisfactory agreement. We close with observations of how, and to what extent, muon diffusion should show up in related compounds.

As previously mentioned, motion of polarons, based on muonium and not unbound muons, was observed in KCl (also in NaCl, GaAs). 
However there is an important difference in the nature of the relaxation in diffusing muonium and diffusing muons. In the former case, the relaxation rate is given by $T_1^{-1} \sim  \delta_{ex}^2 \tau_d/ (1+ \omega_{12}^2 \tau_d^2)$ where $\tau_d^{-1}$ \cite{Kiefl, Kadono, Kadono3} is the diffusion rate (inverse 'time-of-stay' of Mu on one site) and $\delta_{ex}$ is an average electro-nuclear coupling constant $\delta_{ex} \approx \omega_c [n I(I+1)/3]^{1/2}$, $\omega_c$ being the nearest-neighbor atoms nuclear hyperfine parameters, $I$ the atomic spin, and $n$ the number of nearest neighbors. The energy gap $\omega_{12}$ is the smallest intratriplet transition of the Mu spin state related to the (isotropic) contact interaction $\omega_{iso}$ by $\omega_{12} = \omega_{iso}[1+(\Gamma_-/\Gamma_+)x-(1+x^2)^{1/2}]/2$. Here $\Gamma_\pm = (\gamma_{\mu} \pm \gamma_e)/2$, where $\gamma_{\mu}$ ($\gamma_{e}$) stands for the muon (electron) gyromagnetic factor. $x= 2\Gamma_+ B/\omega_{iso}$ is a parameter which varies with the applied longitudinal field $B$ so that measurements at different fields allow one to extract $\delta_{ex}$ and finally the diffusion rate $\tau_d^{-1}$ from the raw data \cite{Kiefl, Kadono,Kadono3,Kadono2}. The value of $\omega_{iso}$ were estimated to be about 4280 MHz \cite{Iso}, while $\delta_{ex}$ is about 50 MHz in KCl (almost temperature independent \cite{Kadono}). Since $\omega_{iso}$ is large, close to the minimum of $\tau_d^{-1}$ as function of temperature, we have $\omega_{12} \tau_d \gg 1$, so that the relaxation rate is found to be roughly proportional to the diffusion rate: $T_1^{-1} \sim (\delta_{ex}^2/ \omega_{12}^2) \tau_d^{-1}$.
 
\begin{figure}
\includegraphics[scale=0.46]{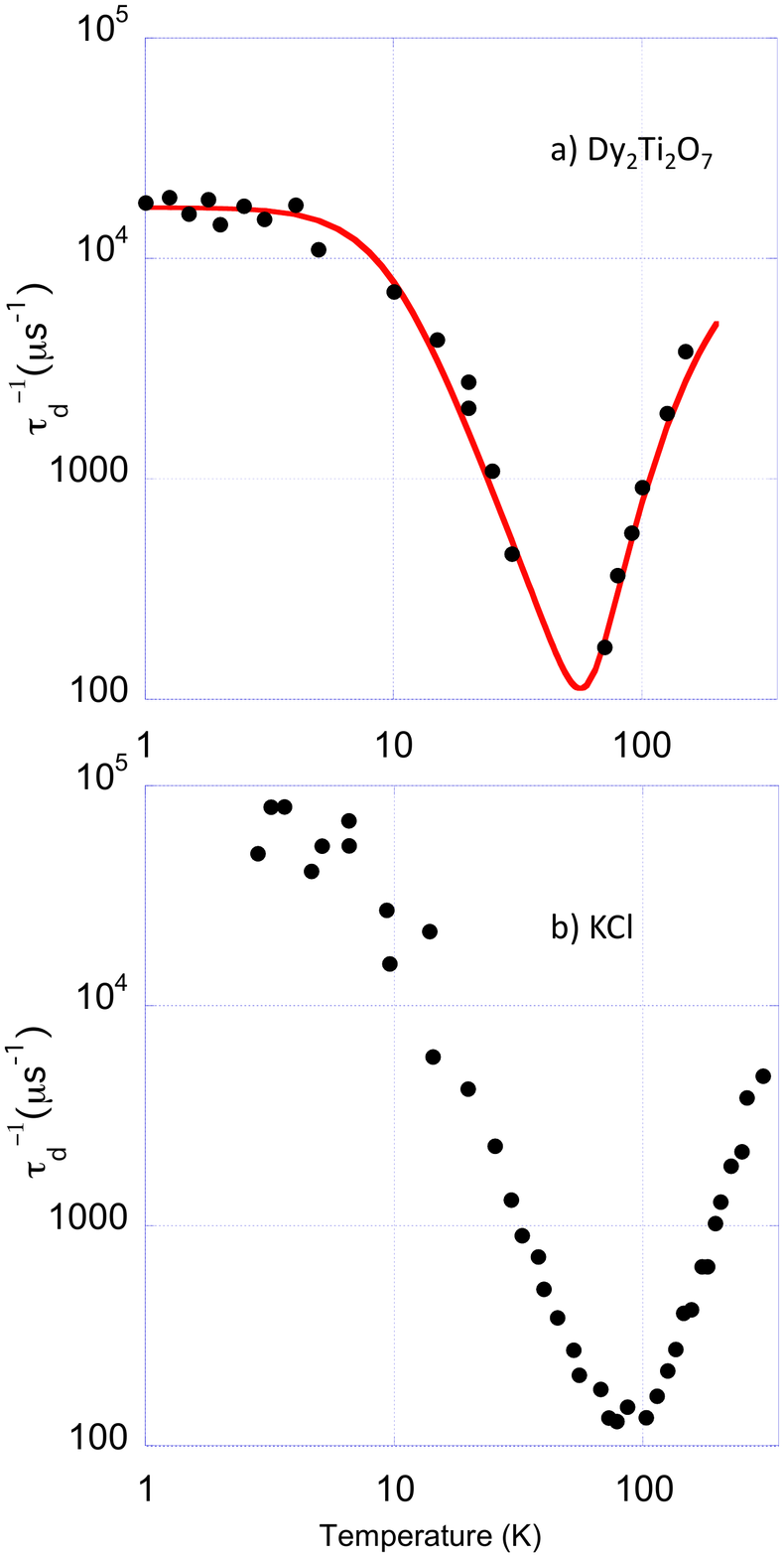}
\caption{\label{fig:fit} Comparison of the diffusion rate (inverse correlation time) $\tau_d^{-1}$ of (a) DTO extracted from the data of Ref. \cite{Dunsiger} by using $\tau_d^{-1}=\omega_0^2 T_1$, where we have fixed $\omega_0 = 1$ Tesla and (b) KCl taken from Ref.~\cite{Kagan4}. The original KCl data may be found in Ref.~\cite{Kiefl,Kadono}. The fit in panel (a) follows the formula (\ref{fit}) with $\alpha=3$, $\nu_0=1/\tau_0= 1.7 \cdot 10^{10}$ $s^{-1}$, $E_a=450$ K, $\nu_1= 6.8 \cdot 10^{11}$s$^{-1}$K$^{1/2}$ and $\nu_2=3.5 \cdot 10^5$ s$^{-1}$. $\Theta_D=350$ K is the estimated Debye temperature (value taken from \cite{Hiroi}).}
\end{figure}

By contrast, in the case of a diffusive muon, Kondo \cite{Kondo2} found that $T_1^{-1}= \omega_0^2 \tau_d$ indicating a motional narrowing decay provided $\omega_0 \tau_d \ll 1$ (which is the case presently). Thus, for Mu the relaxation rate is (roughly) \textit{proportional} to the diffusion rate, whereas for a muon it is \textit{inversely proportional}. $\omega_0$ is the second moment of the distribution of fields due to the magnetic ions in the compound. In spin ice it is generally considered to be large, and we have fixed $\omega_0 \sim 1$ T throughout this paper \cite{Dunsiger, Sala}. 
Most importantly, for short times, the asymmetry function $A(t)$ within this mechanism is a Gaussian function of time \cite{Kondo2} $A(t) \sim \exp (-\omega_0^2 t^2)$, so that at the minimum time of measurement, which is about $10^{-8}$s, with $A(t= 10^{-8} \text{s}) \approx 0.15$ already strongly reduced from the fully polarized limit, as is observed experimentally.
From the relaxation rate data as a function of the temperature (Fig.3 in Ref.~\cite{Dunsiger}), we have extracted $1/\tau_d = \omega_0^2 T_1$ which is the diffusion rate of the muon assuming our basic hypothesis. This is represented in our Fig.~\ref{fig:fit}, together with the diffusion rate previously measured in KCl \cite{Kiefl, Kadono,Kagan4}. We see that the resemblance is quite appealing : the diffusion rate is of the same order of magnitude, its range of variation is about $2$-$3$ orders of magnitude in both cases, and finally the cross-over temperature (minimum of the curve) is more or less the same ($50$ K for spin ice, $70$ K for KCl). An important point to note is that the value of $\tau_d^{-1}$ at the true minimum (around $50$ K) in the case of spin ice cannot be measured : this is due to the fact that it would lead to values of $T_1^{-1}$ which are beyond the experimental limit of $\sim 10^{-8}$ s~\cite{Dunsiger}. 

Now, we are left with the problem of understanding the polaronic diffusion in the substance as a function of temperature. This problem, which belongs to the general topic of the quantum diffusion of heavy particles in solids, has been widely studied in the \cite{Holstein,Kondo1, Kagan1, Petzinger, Klinger1, Leggett1, Kagan2, Kagan3, Storchak,Stamp}. All microscopic theories agree that polaronic diffusion occurs in three temperature regimes (see Fig.~\ref{fig:colour}). First, the very low temperature regime (lower than about $7$ K in our case) is characterized by a diffusion rate independent of temperature: this is the tunneling regime of the polaron, where the muon surrounded by its phonon cloud tunnels through the lattice as a whole as if it were a rigid free particle. This is a coherent band motion. The high temperature regime (above $T^*=50$ K in our case) is a thermally activated one characterized by an excitation energy. In this regime the polaron jumps from site to site. Finally in the intermediate regime, the motion is still incoherent but its coherence progressively increases as the temperature decreases. 

In the spirit of works considering quantum diffusion in KCl (following Kiefl \cite{Kiefl} and Kagan and Prokofiev \cite{Kagan4}), we have fitted the diffusion rate with the following formula:
\begin{equation}
\label{fit}
\frac{1}{\tau_d} = \frac{\nu_1}{ \sqrt{T}} e^{-E_a/T}+ \frac{\nu_0}{1+(\nu_0/\nu_2) \left(T/ \Theta_D \right)^\alpha}
\end{equation}
The first term dominates at high $T > T^*$ and corresponds to the usual activated regime of the polaron motion \cite{Holstein,Stamp}. The second term interpolates the very low temperature regime with the intermediate one \cite{Kagan4}. In the intermediate temperature range, $1/ \tau_d \sim \nu_2 (T/\Theta_D)^{-\alpha}$ \cite{Kiefl}, whereas at low temperature $1/ \tau_d \sim \nu_0$ corresponds to the tunneling regime.

Our fit is shown in Fig.~\ref{fig:fit}, with the values of the different parameters used. The parameters appropriate for acoustic phonons give $\alpha=3$ (see supplemental material) and $\Theta_D=350$ K as quoted in the literature \cite{Hiroi}. This leaves four fitting parameters $\nu_0=1.7 \cdot 10^{10}$s$^{-1}$, $\nu_1= 6.8 \cdot 10^{11}$s$^{-1}$K$^{1/2}$, $\nu_2= 3.5 \cdot 10^{5}$s$^{-1}$ and $E_a=450$ K. By comparison, in case of KCl, the different parameters were found to be \cite{Kiefl}: $E_a \sim 390$ K, $\nu_0 \sim 5 \cdot 10^{10}$s$^{-1}$, $\nu_2 \sim 1.3 \cdot 10^{6}$s$^{-1}$. The parameter $\nu_1$ cannot be directly compared because in \cite{Kiefl} a pure exponential law $\nu_1 \exp(-E_a/T)$ was used, while our $(\nu_1/\sqrt{T}) \exp(-E_a/T)$ law is more appropriate for a polaron. At $T=100$ K, we find $\nu_1/\sqrt{T}=6.8 \cdot 10^{10}$s$^{-1}$, while the value in KCl is $8.3 \cdot 10^{9}$s$^{-1}$. The different parameters are roughly of the same order in DTO and KCl. Although Mu is a neutral particle, it is is a \textit{composite} one (muon + electron), whose parts interact differently with both acoustic and optical phonons (muons and electrons have quite different masses, and thus different effective particle-phonon interaction). We do not discuss this point any further, but we think that both types of phonon may also be important for the diffusion of Mu in KCl and that is the origin of the similarity observed between the (possible) diffusion of muon in DTO and of Mu in KCl. We note however that to our knowledge, the composite nature of the Mu has never been taken into account for its diffusion in any microscopic model.

Equation~\ref{fit} contains four fitting parameters. These can be related to the parameters of a microscopic model (see Supplemental Material for details) \cite{Holstein,Petzinger} to provide estimates of their values, which we find to be entirely consistent with the fit we have extracted from the experimental data. To summarize the content of the model, we have three distinct temperature regimes. At high temperatures, there is a thermally activated hopping regime, whereas at very low temperature there is a coherent tunneling regime. Between both, there is a third intermediate temperature regime for the hopping rate which exhibits a $\sim T^{-3}$ law. 

At high temperatures, the hopping rate is:
\begin{equation}
\frac{1}{\tau_d} \sim \frac{1}{\hbar\sqrt{ E_a (k_BT)}} \Delta_0^2 e^{-E_a/k_BT}.
\end{equation}
Here $E_{a}$ is the polaron activation energy which is of the order of the typical phonon energy \cite{Holstein,Petzinger}. $\Delta_0$ is the tight-binding hopping term of a muon, related to the total bandwidth $W\sim Z\Delta_{0}$ ($Z$ is the coordination number of the lattice). Since typical bandwidth for the electron is of the order of a few eV, the bare muonic bandwidth may be estimated to be of the order of $(m_e/m_{\mu}) \approx 1/200$ times the typical electronic bandwidth. That gives $\Delta_0 \sim 10^{-3}$ eV. The coefficient of $(1/\sqrt{T})\exp(-E_{a}/k_{B}T)$ is just $\nu_1$. On the basis of our estimated $\Delta_0$ and the fitted activation energy $E_{a}\sim 450$ K, we find that $\nu_{1}\sim 10^{11}$ s$^{-1}$K$^{1/2}$, which is of the same order of magnitude than the corresponding fitting parameter.

In the intermediate temperature regime ($T<T^*$), the hopping rate becomes
\begin{equation}
\frac{1}{\tau_d} \sim \frac{\tilde \Delta_0^2}{\hbar (k_B \Theta_D)} \left( \frac{T}{\Theta_D}\right)^{-3}.
\end{equation}
This fixes the coefficient $\nu_2$ and $\alpha$. Then $\nu_2 =  \tilde \Delta_0^2/\hbar (k_B \Theta_D)$.  In this formula, $\tilde \Delta_0$ is the polaron bandwidth which is strongly reduced from the bare muonic bandwidth by a reduction factor $\exp(-S)$: $\tilde \Delta_0 = e^{-S} \Delta_0$. Usually, $e^{-S}$ is estimated to be $10^{-2}-10^{-4}$ (see Ref.~\cite{Holstein}). By taking $10^{-3}$, we obtain $\nu_{2}\sim 10^{5}$ s$^{-1}$ in agrement with the fitting parameter.

Finally, at low temperatures, there is a coherent regime for which 
\begin{equation}
 \frac{1}{\tau_d} \sim \tilde \Delta_0/\hbar.
 \end{equation}
This corresponds to the fitting parameter $\nu_{0}$. Since we have already estimated $\tilde\Delta_0 \sim 10^{-3} \Delta_0$ we may directly check that it gives $\nu_0 \sim 10^{10}$ s$^{-1}$, also in agreement with its fitted value.
 
Our main conclusion is that the measurements of Dunsiger {\it et al.} \cite{Dunsiger} appear to be completely compatible with the observation of quantum diffusion of a muonic polaron in its three regimes of temperature. Indeed, the magnetism in this particular material apparently offers a beautiful probe of the motional crossover phenomenon as the temperature is varied. We also propose an experimental check of this basic hypothesis \cite{Barbara}: one may carry out the same $\mu$SR experiment on \textit{non-magnetic} compounds within the same family as DTO such as Y$_2$Ti$_2$O$_7$ and Lu$_2$Ti$_2$O$_7$. If the mechanism we put forward in this paper is correct, muon spin relaxation should also occur due to muon diffusion, but now through the local fields of the nuclear spins. (If no nuclear spins were present, muon diffusion would become unobservable in $\mu$SR.) If the muons are localized, there would \textit{a priori} be no possibility of relaxation anymore (since the nuclear-spins have very large relaxation times), except by the Kubo-Toyabe mechanism which, on its own, would generate an asymptotic $1/3$ tail in the asymmetry. We have found only one example of such an experiment in the literature \cite{Dunsigerthesis}, in which a relaxation plateau with a relaxation rate of about $1$ $\mu$s$^{-1}$ was indeed measured. This is consistent with the basic hypothesis that the muon spin relaxation comes from the diffusion of the muon. However, we think that further investigations are needed to clarify this very interesting physics.

As a final remark, we note that constant relaxation rates at low temperatures have been almost systematically observed in other frustrated magnets. Typically, the constant relaxation rate gives way, at higher temperatures, to clear signs of the intrinsic magnetism in the material. For example, one often observes signs of magnetic phase transitions using $\mu$SR in agreement with other experimental probes. While quantum diffusion may offer an explanation for the low temperature relaxation plateaux beyond the case of DTO, there is often no sign of the two higher temperature regimes discussed in this paper. However, two time scales will be relevant to the experimentally probed relaxation rate: the diffusion rate of the polaronic muon $\tau_d$ and $\tau_f$, the characteristic fluctuation time for the magnetic spins. Both quantities vary with temperature. When $\tau_f \ll \tau_d$, the muon relaxation is mainly driven by the spin fluctuations, and the muon diffusion may be ignored in that case. On the contrary, when $\tau_d \ll \tau_f$, the muon diffusion drives the relaxation. This is probably what happens at low temperature in the pyrochlore compounds, which could explain the systematically observed plateau at very low temperature. In the spin ice material considered in this article, our scenario supposes that the spin dynamics gives a subdominant contribution to the muon relaxation at low temperature and that $\tau_d < \tau_f$ for all temperatures. Finally, we note that a general theory which includes effects of both $\tau_d$ and $\tau_f$ simultaneously seems to be absent in the literature \cite{McClarty}, although McMullen and Zaremba \cite{Zaremba} partly discussed this case.

We would like to thank the following physicists for very useful discussions : B. Barbara, S. Dunsiger, A. Keren, and T. Uemura. We also thank C. Castelnovo, M. Gingras and S. Sondhi for these and collaborations on much related work.

\appendix

\section{Supplementary Material}

To represent muon diffusion through the lattice, we start with the following tight-binding Hamiltonian \cite{Petzinger}:
\begin{eqnarray}
H &=& \frac{1}{2} \sum_{\mathbf{q} \lambda} \omega_{\mathbf{q} \lambda} \left( \vert p_{\mathbf{q}\lambda} \vert^2 + \vert u_{\mathbf{q} \lambda} \vert^2 \right) - \Delta_0 \sum_{\mathbf{n},\mathbf{\delta}} \left( c_{\mathbf{n}+\mathbf{\delta}}^{+} c_{\mathbf{n}}^{}+h.c. \right) \nonumber \\
&-& \frac{1}{\sqrt{N}} \sum_{\mathbf{q} \lambda} \sum_{\mathbf{n}} \gamma_{\mathbf{q}, \lambda} u_{\mathbf{q} \lambda} e^{i \mathbf{q} \cdot \mathbf{R}_{\mathbf{n}}} c_{\mathbf{n}}^{+} c_{\mathbf{n}}^{}.
\end{eqnarray}
The first term is the phonon Hamiltonian and the second term corresponds to the band motion of the muon ($\mathbf{\delta}$ is a vector connecting two neighboring sites of the lattice). Finally, the last term is the muon-phonon coupling. Since the muon is a charged particle, it interacts with both the optical and acoustic phonons ($\lambda$ is a phonon branch index). For the sake of simplicity, we also assume in the following that the muon sites form a simple three-dimensional cubic lattice. Let us now introduce the lattice operators:
\begin{eqnarray}
u_{\mathbf{q},\lambda} &=& (a^{}_{\mathbf{q}, \lambda}+a^+_{-\mathbf{q}, \lambda}) /\sqrt{2}\nonumber \\
p_{\mathbf{q},\lambda} &=& (a^{}_{-\mathbf{q}, \lambda}-a^+_{\mathbf{q}, \lambda}) /\sqrt{2}i.
\end{eqnarray}
As is usual in the theory of polarons, we also introduce the following operators:
\begin{equation}
\phi_{\mathbf{n}} = \frac{1}{\sqrt{N}} \sum_{q \lambda} \frac{\gamma_{\mathbf{q},\lambda}^*}{\omega_{q \lambda}} p_{\mathbf{q} \lambda} e^{i \mathbf{q} \cdot \mathbf{R}_{\mathbf{n}}},
\end{equation}
\begin{equation}
\phi=\sum_{\mathbf{n}} \phi_{\mathbf{n}} c_{\mathbf{n}}^+c_{\mathbf{n}}^{},
\end{equation}
and we perform the Lang-Firsov unitary transformation on the Hamiltonian (1) \cite{Langfirsov,Petzinger},
\begin{equation}
e^{i \phi} H e^{-i \phi}
\end{equation}
The resulting transformed Hamiltonian is:
\begin{eqnarray}
H' &=& \frac{1}{2} \sum_{\mathbf{q} \lambda} \omega_{\mathbf{q} \lambda} \left( \vert p_{\mathbf{q}\lambda} \vert^2 + \vert u_{\mathbf{q} \lambda} \vert^2 \right)- E_a\sum_{\mathbf{n}} c_{\mathbf{n}}^+c_{\mathbf{n}}^{} \nonumber \\
&-& \Delta_0 \sum_{\mathbf{n}, \mathbf{\delta}} e^{i(\phi_{\mathbf{n}}-\phi_{\mathbf{n}+\mathbf{\delta}})} c_{\mathbf{n}+\mathbf{\delta}}^+c_{\mathbf{n}}^{} 
\end{eqnarray}
with
\begin{equation}
E_a = \frac{1}{2N} \sum_{\mathbf{q} \lambda} \frac{\vert  \gamma_{\mathbf{q}\lambda} \vert^2}{\omega_{\mathbf{q} \lambda}}.
\end{equation}
Let us write the last term of the expression (6) in a different form:
\begin{eqnarray}
\Delta_0 \sum_{\mathbf{n}, \mathbf{\delta}} e^{i(\phi_{\mathbf{n}}-\phi_{\mathbf{n}+\mathbf{\delta}})} c_{\mathbf{n}+\mathbf{\delta}}^+c_{\mathbf{n}}^{} &=& \Delta_0 \sum_{\mathbf{n}, \mathbf{\delta}} \big< e^{i(\phi_{\mathbf{n}}-\phi_{\mathbf{n}+\mathbf{\delta}})} \big>_{ph}c_{\mathbf{n}+\mathbf{\delta}}^+c_{\mathbf{n}}^{} \nonumber \\
&+& \Delta_0 \sum_{\mathbf{n}, \mathbf{\delta}} \Big[ e^{i(\phi_{\mathbf{n}}-\phi_{\mathbf{n}+\mathbf{\delta}})}- \big< e^{i(\phi_{\mathbf{n}}-\phi_{\mathbf{n}+\mathbf{\delta}})} \big>_{ph} \Big] c_{\mathbf{n}+\mathbf{\delta}}^+c_{\mathbf{n}}^{}.
\end{eqnarray}
The expression $\big< A \big>_{ph}$ means that we take the mean value of the operator $A$ on the phonon states $\big \vert \dots N_{\mathbf{k}\lambda} \dots N_{\mathbf{k}' \lambda} \dots \big>$, where $N_{\mathbf{k} \lambda}$ are the phonon occupation numbers,
\begin{equation}
N_{\mathbf{k} \lambda} = \frac{1}{e^{-\beta \omega_{\mathbf{k} \lambda}}-1}
\end{equation}
The expression (8) defines \textit{two different channels} for the polaron motion from one site to the neighboring site: a coherent channel where the phonons occupations numbers do not change during the motion (diagonal term in (8)), and an incoherent channel (jump motion) where changes in the phonon occupations numbers during the motion are allowed. At very low temperatures, the second process can be neglected, whereas it becomes predominant at higher temperatures. The diagonal term can be calculated and gives:
\begin{equation}
\big< e^{i(\phi_{\mathbf{n}}-\phi_{\mathbf{n}+\mathbf{\delta}})} \big>_{ph} = e^{-S(T)}
\end{equation}
\begin{equation}
 S(T) = \frac{1}{N} \sum_{\mathbf{q} \lambda} \frac{\vert  \gamma_{\mathbf{q}\lambda} \vert^2}{\omega_{\mathbf{q} \lambda}^2} \coth{\left(\beta \omega_{\mathbf{q} \lambda}/2 \right)} \sin^2{\left[ \left(\mathbf{q} \cdot \mathbf{\delta} \right)/2 \right]}.
\end{equation}

\bigskip
\bigskip
\textit{Coherent motion}. Let us first examine the coherent motion at very low temperatures. In this regime, all non-diagonal phonon transitions are negligible so that the relevant effective Hamiltonian for the polaron motion can be reduced to:
\begin{eqnarray}
H_{coh} &=& -E_a \sum_{\mathbf{n}} c_{\mathbf{n}}^+c_{\mathbf{n}}^{} - \Delta_0 e^{-S(T)} \sum_{\mathbf{n}, \mathbf{\delta}}
c_{\mathbf{n}+\mathbf{\delta}}^+c_{\mathbf{n}}^{} .
\end{eqnarray}
The eigenstates are $\epsilon(\mathbf{k})=-E_a-2 \tilde \Delta_0 \left[\cos(k_x)+\cos(k_y)+\cos(k_z) \right]$, with $\tilde \Delta_0=\Delta_0 e^{-S(T)}$ and the Green's function is
\begin{equation}
G_{\mathbf{q}}(t)= \frac{1}{N} \sum_{\mathbf{k}} e^{i(\epsilon_{\mathbf{k}}-\epsilon_{\mathbf{k}+\mathbf{q}})t}.
\end{equation}
From the above expression we can calculate the mean time of stay of the muon on one site, i.e. its inverse tunneling rate:
\begin{equation}
\tau = \frac{\hbar}{N^2} \int_0^{\infty} \left( \sum_{\mathbf{k},\mathbf{q}} e^{i(\epsilon_{\mathbf{k}}-\epsilon_{\mathbf{k}+\mathbf{q}})t}\right) dt \approx \frac{\hbar}{2 \sqrt{2} \tilde \Delta_0} \sim \frac{\hbar}{\tilde \Delta_0}.
\end{equation}
The expression (11) for $S$ thus defines the polaron reduction factor of the bandwidth in the coherent regime.

\bigskip
\bigskip
\textit{Incoherent hopping rate:} At intermediate and large temperatures, the phonon occupation numbers $N_k$ increase so that the second \textit{channel} for the diffusion becomes predominant. At these temperatures, the delocalized states $\mathbf{k}$ lose their meaning and the effective Hamiltonian for the polaron diffusion is reduced to:
\begin{equation}
H_{incoh} = -E_a \sum_{\mathbf{n}} c_{\mathbf{n}}^+c_{\mathbf{n}}^{}- \Delta_0 \sum_{\mathbf{n}, \mathbf{\delta}} \Big[ e^{i(\phi_{\mathbf{n}}-\phi_{\mathbf{n}+\mathbf{\delta}})}- \big< e^{i(\phi_{\mathbf{n}}-\phi_{\mathbf{n}+\mathbf{\delta}})} \big>_{ph} \Big] c_{\mathbf{n}+\mathbf{\delta}}^+c_{\mathbf{n}}^{}.
\end{equation}
From the second term, we can calculate a transition probability $W(\mathbf{n} \rightarrow \mathbf{n}\pm \mathbf{\delta})$ to second order in perturbation theory, and the corresponding jump (or hopping) rate. Holstein carried out this calculation and got the following hopping rate \cite{Holstein}:
\begin{eqnarray}
\frac{1}{\tau} &=& \frac{\Delta_0^2}{\hbar} \pi^{1/2} \left[ \sum_{\mathbf{k} \lambda}^{}  \vert \gamma_{\mathbf{k} \lambda} \vert^2 \sin^2 \left[(\mathbf{k} \cdot \delta)/2\right] \text{csch} \left(\beta \hbar \omega_{\mathbf{k} \lambda}/2 \right) \right]^{-1/2}  \nonumber \\
 &\times& \exp \left[ -2 \sum_{\mathbf{k} \lambda}  \frac{\vert \gamma_{\mathbf{k} \lambda} \vert^2}{\omega_{\mathbf{k} \lambda}^2} \sin^2 \left[ (\mathbf{k} \cdot \delta)/2\right] \tanh \left(\beta \omega_{\mathbf{k} \lambda}/4 \right)\right].
\end{eqnarray}
(we recall that $\text{csch}(x) = \sinh (x)^{-1}$).

In the high temperature limit, i.e. $T \gg \omega/2$, we have $\tanh \left(\beta \omega/4 \right) \sim \beta \omega/4$ and $\text{csch} \left(\beta \omega/2 \right) \sim 2/ \beta \omega$, and using the definition (7) of $E_a$, (16) becomes
\begin{equation}
\frac{1}{\tau} \sim \frac{\pi^{1/2}}{\hbar \sqrt{E_a (k_BT)}} \Delta_0^2 e^{-E_a/k_BT},
\end{equation}
which corresponds to our expression (4) in the paper. This is the thermally activated regime.

In the intermediate temperature regime which should nevertheless be a low temperature with respect to the phonon frequencies, $\tanh (\beta \omega) \sim 1$ and $\text{csch} (\beta \omega) \sim \exp (-\beta \omega)$ so that (16) becomes by using the definition (11),
\begin{equation}
\frac{1}{\tau} = \frac{\tilde \Delta_0^2}{\hbar} \pi^{1/2} \left[ \sum_{\mathbf{k} \lambda}^{}  \vert \gamma_{\mathbf{k} \lambda} \vert^2 \sin^2 \left[(\mathbf{k} \cdot \delta)/2\right] e^{- \left(\beta \omega_{\mathbf{k} \lambda}/2 \right)} \right]^{-1/2}.
\end{equation}
To evaluate this expression, we must now separate the respective role of the optical and acoustical branches. The optical branches give the main contribution to the reduction factor $S(T)$ \cite{Holstein}, whereas the contribution to this factor coming from the acoustical branches is much less \cite{Stamp}. However in this intermediate temperature regime, the acoustic branches play a major role in the polaron motion. The reason is that the exponential factors in (18) vanish for all optical phonon frequencies. Consequently, we may only consider the acoustic phonons in the summation over the different branches in (18). Let us now calculate the factor
\begin{equation}
\sum_{\mathbf{k}_{acc.}}^{}  \vert \gamma_{\mathbf{k}_{acc.}} \vert^2 \sin^2 \left[(\mathbf{k}_{acc.} \cdot \delta)/2\right] e^{- \left(\beta \omega_{\mathbf{k}_{acc.}}/2 \right)}
\end{equation} 
We adopt the Debye approximation for the acoustic phonons, and following Petzinger \cite{Petzinger}, we take for all longitudinal modes
\begin{eqnarray}
E_{acc.} &\approx& \vert \gamma_{\mathbf{k}} \vert^2/2 \omega_{\mathbf{k}}, \nonumber \\
\sin^2 \left[(\mathbf{k} \cdot \delta)/2\right] &\approx& \omega_{\mathbf{k}}^2/\omega_D^2,
\end{eqnarray}
where $\omega_D$ is the Debye frequency ($\hbar \omega_D = k_B \Theta_D = k_D c/\hbar$, with $\Theta_D$ the Debye temperature, $c$ the sound velocity and $k_D \approx \pi/a$ the Debye wave vector). By introducing the phonon density of states in the Debye approximation $g_D(\omega) = \left[3/(2 \pi^2)\right] \omega^2/\omega_D^3$, and replacing the summation in (19) by an integral, we easily get
\begin{equation}
\frac{1}{\tau} \sim \frac{\tilde \Delta_0^2}{\hbar \sqrt{\left( k_B \Theta_D \right) E_{acc.} }} \left( \frac{T}{\Theta_D} \right)^{-3},
\end{equation}
where all the numerical factors together have been evaluated to be of the order of unity. The precise value of $E_{acc.}$ is not known, but is necessarily of the order of $k_B \Theta_D$, so that we have finally
\begin{equation}
\frac{1}{\tau} \sim \frac{\tilde \Delta_0^2}{\hbar ( k_B \Theta_D)} \left( \frac{T}{\Theta_D}\right)^{-3},
\end{equation}
which is the formula (3) of our paper, which is applied in the intermediate regime.

Finally, we have adopted the simplest theory of the quantum diffusion of a muonic polaron to justify our fit, and it appears quite satisfying. However, other possibilities such as two-phonon interaction processes and/or a fluctuation preparation barrier \cite{Storchak} should be also examined. Possible Umklapp processes, when several muon sites per unit cell exist, should also be studied. We reserve all these technical discussions for a further publication.


\begin{thebibliography}{99}
%
\bibitem{Dunsiger} S.R. Dunsiger \textit{et al.}, Phys. Rev. Lett. \textbf{107}, 207207 (2011).
%
\bibitem{Bert} F. Bert \textit{et al.}, Physica B {\bf 374}, 134 (2006).
\bibitem{Uemura} Y. J. Uemura \textit{et al.}, Phys. Rev. Lett. {\bf 73} 3306 (1994).
\bibitem{Fukaya} A. Fukaya \textit{et al.}, Phys. Rev. Lett. {\bf 91} 207603 (2003).
\bibitem{Keren2} A. Keren \textit{et al.}, Phys. Rev. Lett. {\bf 84} 3450 (2000).
\bibitem{Hodges} J. A. Hodges \textit{et al.}, Phys. Rev. Lett. {\bf 88} 077204 (2002).
\bibitem{Gardner} J. S. Gardner \textit{et al.}, Phys. Rev. Lett. {\bf 82} 1012 (1999).
\bibitem{TTOMuon} A. Keren \textit{et al.}, Phys. Rev. Lett. {\bf 92} 107204 (2004).
\bibitem{GSO} P. A. McClarty {\it et al.} J. Phys.:Condens. Matter {\bf 23} 164216 (2011). 
\bibitem{Bramwell1} S. T. Bramwell \textit{et al.}, Nature, \textbf{461}, 956 (2009).
%
\bibitem{Bramwell2} S. T. Bramwell and S. R. Giblin, arXiv:1111.4168 (2011).
%
\bibitem{Blundell}  S. J. Blundell, arXiv:1111.3657 (2011).
%
\bibitem{Sala} G. Sala \textit{et al.} arXiv:1112.3363 (2011).
%
\bibitem{LagoBlundell} J. Lago, S. J. Blundell and C. Baines {\bf 19} 326210 (2007).
%
\bibitem{Snyder} J. Snyder \textit{et al.}, Phys. Rev. B \textbf{70}, 184431 (2004).
\bibitem{Matsuhira} K. Matsuhira \textit{et al.}, J. Phys.: Condens. Matter {\bf 12} L649 (2000).
\bibitem{Quilliam} J. P. Quilliam {\it et al.} arXiv:1102.1703
\bibitem{Matsuhira2}  K. Matsuhira \textit{et al.}, J. Phys. Soc. Jpn. {\bf 80} 123711 (2011).
\bibitem{Ehlers} G. Ehlers \textit{et al.}, J. Phys.: Condens. Matter {\bf 16} S635 (2004).
\bibitem{Ehlers2} G. Ehlers \textit{et al.}, J. Phys.: Condens. Matter {\bf 15} L9 (2003).
\bibitem{Clancy} J. P. Clancy \textit{et al.}, Phys. Rev. B {\bf 79} 014408 (2009).
\bibitem{Orendac} M. Orend\'{a}\`{c} {\it et al.}, Phys. Rev. B {\bf 75} 104425 (2007).
\bibitem{Sutter} J. P. Sutter \textit{et al.}, Phys. Rev. B \textbf{75}, 140402(R) (2007).
\bibitem{Kitagawa} K. Kitagawa, R. Higashinaka, K. Ishida, Y. Maeno, and M. Takigawa, Phys. Rev. B {\bf 77} 214403 (2008).
\bibitem{Jaubert} L. Jaubert and P. Holdsworth, Nature Phys. {\bf 5} 258 (2009).
%
\bibitem{Kagan4} Y. Kagan and N. V. Prokofiev, Phys. Lett. A \textbf{150}, 320 (1990).
\bibitem{Kiefl} R.F. Kiefl \textit{et al.}  Phys. Rev. Lett. \textbf{62}, 792 (1989).
%
\bibitem{Kadono} R. Kadono \textit{et al.}, Phys. Rev. Lett. \textbf{64}, 665 (1990).
%
\bibitem{Kadono3} R. Kadono, R. F. Kiefl, W. A. MacFarlane, and S. R. Dunsiger, Phys. Rev. B \textbf{53}, 3177 (1996). 
%
\bibitem{Kadono2} R. Kadono \textit{et al.}, Hyp. Int. \textbf{64}, 635 (1990).
%
\bibitem{Storchak} V. G. Storchak, N.V. Prokofiev, Rev. Mod. Phys. \textbf{70}, 929 (1998).
%
\bibitem{Bi} C.Z. Bi \textit{et al.}, J.Phys.: Condens. Matter \textbf{17}, 5225 (2005).
\bibitem{Iso} R.F. Kiefl {\it et al.}, Phys. Rev. Lett. \textbf{53}, 90 (1984).
\bibitem{Kondo2} J. Kondo, Hyp. Inter. \textit{105}, 203 (1997).
%
\bibitem{Hiroi} Z. Hiroi \textit{et al.} J. Phys. Soc. Jpn. \textbf{72}, 411 (2003). Our fit is insensitive to the considerable reported variation of $\Theta_D$ \cite{Klemke}.
\bibitem{Klemke} B. Klemke {\it et al.} J. Low Temp. Phys. {\bf 163} 345 (2011).
%
\bibitem{Holstein} T. Holstein, Annals of Physics \textbf{8}, 343 (1957); ibid. \textbf{8}, 325 (1957).
%
\bibitem{Kondo1} J. Kondo, Hyp. Inter. \textit{31}, 117 (1986).
%
\bibitem{Kagan1} Y. Kagan, M. I. Klinger, J. Phys. C: Sol. Stat. Phys. \textbf{7}, 2791 (1974).
%
\bibitem{Petzinger} K. G. Petzinger, Phys. Rev. B \textbf{26}, 6530 (1982).
%
\bibitem{Klinger1} M. I. Klinger, Phys. Rep. \textbf{94}, 183 (1983).
Å%
\bibitem{Leggett1} A. J. Leggett \textit{et al.}, Rev. Mod. Phys. \textbf{59}, 1 (1987).
%
\bibitem{Kagan2} Y. Kagan, J. Low Temp. Phys. \textbf{87}, 525 (1992).
%
\bibitem{Kagan3} Y. Kagan and N. V. Prokofiev, in \textit{Quantum Tunnelling in Condensed Media}, ed. Y. Kagan and A. J. Leggett, p.37-143, Elsevier Science Publishers 1992.
%
\bibitem{Stamp} P. C. E Stamp and Chao Zhang, Phys. Rev. Lett. \textbf{66}, 1902 (1991).
%
\bibitem{Barbara} B. Barbara and P. Qu\'emerais, private communication.
%
\bibitem{Dunsigerthesis} S. R. Dunsiger, \textit{Spin Relaxation in Geometrically Frustrated Pyrochlores}, Ph.D. dissertation, University of British Columbia (2000) ; unpublished.
%
\bibitem{McClarty} P. McClarty \textit{et al.}, in preparation.
%
\bibitem{Zaremba} T. McMullen and E. Zaremba, Phys. Rev. B \textbf{18}, 3026 (1978).
\end{thebibliography}
\end{document}